\documentclass[journal]{IEEEtran}
\usepackage{booktabs}
\usepackage{amssymb}
\usepackage{amsfonts}
\usepackage{bbding}
\usepackage{svg}
\usepackage{subfigure}
\usepackage{amsmath} 
\usepackage[numbers]{natbib}
\usepackage{url}
\usepackage{multirow}
\usepackage{pifont}

\ifCLASSOPTIONcompsoc
 \usepackage[caption=false,font=normalsize,labelfont=sf,textfont=sf]{subfig}
\else
 \usepackage[caption=false,font=footnotesize]{subfig}
\fi
\usepackage{dblfloatfix}
\usepackage{url}
\begin{document}

\def\UrlFont{\em}

\title{LWFNet: Coherent Doppler Wind Lidar-Based Network for Wind Field Retrieval}

\author{Ran Tao,
Chong Wang*, 
Hao Chen*, 
Mingjiao Jia, 
Xiang Shang,
Luoyuan Qu, 
Guoliang Shentu, 
Yanyu Lu, 
Yanfeng Huo,
Lei Bai,
Xianghui Xue and Xiankang Dou
 \thanks{
 Ran Tao and Xianghui Xue are with the University of Science and Technology of China, Hefei 230026, China and Hefei National Laboratory, Hefei 230088, China (e-mail: taoran201@mail.ustc.edu.cn; xuexh@ustc.edu.cn).

 Chong Wang is with the University of Science and Technology of China, Hefei 230026, China (e-mail: wclhy50@ustc.edu.cn).

 Chen Hao and Lei Bai are with the Shanghai Artificial Intelligence Laboratory, Shanghai 200000, China (e-mail: justchenhao@buaa.edu.cn; bailei@pjlab.org.cn).

 Mingjiao Jia, Xiang Shang, Luoyuan Qu, and Guoliang Shentu are with the Shandong Guoyao Quantum Lidar Company Ltd., Jinan 250101, China (email: mjjia@lidarq.com; xshang@lidarq.com; quly@lidarq.com; shentu@lidarq.com).

 Yanyu Lu and Yanfeng Huo are with the Anhui Institute of Meteorological Sciences, Hefei 230031, China (e-mail: ahqxlyy@163.com; huoyanfeng@ahmi.org.cn).

 Xiankang Dou is with the Hefei National Laboratory, Hefei 230088, China (e-mail: dou@ustc.edu.cn).

 Corresponding authors: Chong Wang and Hao Chen.
    }
}

\maketitle

\begin{abstract}

Accurate detection of wind fields within the troposphere is essential for atmospheric dynamics research and plays a crucial role in extreme weather forecasting.
Coherent Doppler wind lidar (CDWL) is widely regarded as the most suitable technique for high spatial and temporal resolution wind field detection.
However, since coherent detection relies heavily on the concentration of aerosol particles, which cause Mie scattering, the received backscattering lidar signal exhibits significantly low intensity at high altitudes.
As a result, conventional methods, such as spectral centroid estimation, often fail to produce credible and accurate wind retrieval results in these regions.
To address this issue, we propose LWFNet, the first Lidar-based Wind Field (WF) retrieval neural Network, built upon Transformer and the Kolmogorov-Arnold network.
Our model is trained solely on targets derived from the traditional wind retrieval algorithm and utilizes radiosonde measurements as the ground truth for test results evaluation.
Experimental results demonstrate that LWFNet not only extends the maximum wind field detection range but also produces more accurate results, exhibiting a level of precision that surpasses the labeled targets.
This phenomenon, which we refer to as \textit{super-accuracy}, is explored by investigating the potential underlying factors that contribute to this intriguing occurrence.
In addition, we compare the performance of LWFNet with other state-of-the-art (SOTA) models, highlighting its superior effectiveness and capability in high-resolution wind retrieval.
LWFNet demonstrates remarkable performance in lidar-based wind field retrieval, setting a benchmark for future research and advancing the development of deep learning models in this domain.

\end{abstract}


\section{Introduction}

\IEEEPARstart{T}{he} detection of wind fields plays a pivotal role across various domains, including wind power generation \cite{gonzalez2024offshore, harrison2024artificial, zhang2014review}, wind shear alerts \cite{chan2012application, gao2024interpretable, hon2014application, nechaj2019monitoring} and extreme weather forecasting \cite{dalto2015deep, liu2015emd, klotzbach2019seasonal}. 
Precise wind monitoring enables more effective responses to extreme weather events and helps mitigate their impact on transportation infrastructure \cite{zhao2023hybrid, abbasi2015impact}.
At present, multiple remote sensing techniques have been employed and utilized for wind detection, such as ground-based lidar systems \cite{reitebuch2012windlidar} and global navigation satellite system (GNSS) \cite{rodriguez2012airborneGNSS}.
Wind lidar (direct detection Doppler wind lidar (DDWL) and coherent Doppler wind lidar (CDWL)) measures wind fields by capturing line-of-sight (LOS) velocities from various perspectives using a scanning mode, while the latter produces GNSS-reflectometry (GNSS-R) observations for monitoring ocean surface wind speed (OSWS).

CDWL measures wind speed by emitting laser pulses into the atmosphere and detecting the Doppler frequency shift induced by Mie scattering from moving aerosol particles. 
The Doppler-shifted backscattered signals are typically processed using the fast Fourier transform (FFT), which generates the received lidar spectrum signal.
CDWL has found extensive application in tropospheric wind monitoring, demonstrating the capability to achieve meter-scale spatial precision and sub-second temporal resolution \cite{meter_cdwl}.
Unlike ground-based lidar, GNSS-R is capable of retrieving global ocean wind speeds, albeit at a coarse scale (e.g., 25 km). 
The delay-Doppler map (DDM) of GNSS-R observables, along with the corresponding bistatic radar cross-section (BRCS), reveals the Doppler shift of scattered signals caused by ocean surface winds.
The normalized BRCS and leading edge slope are then used for ocean wind speed retrieval, but its accuracy is highly dependent on the calibration process, which is influenced by numerous factors \cite{said2018assessment, gleason2021characterizing}.

OSWS retrieval methods based on GNSS-R have been extensively studied. 
Geophysical mode functions (GMFs), which are conventional ocean speed retrieval methods, encounter challenges in identifying and accounting for all the relevant parameters involved in the retrieval task \cite{du2024deep}.
As a result, deep learning, with its flexibility in processing multi-source data, has been increasingly utilized by researchers for OSWS retrieval.
Multiple neural networks architectures have been implemented, including fully connected network (FCN) \cite{chu2020multimodal, reynolds2020wind}, convolutional neural network (CNN) \cite{asgarimehr2022gnss, guo2022information, bu2023glows} and Transformer \cite{zhao2023ddm}, demonstrating superior performance compared to traditional retrieval algorithms (e.g., minimum variance estimator \cite{clarizia2014spaceborne}).

In order to study wind field characteristics within the troposphere, particularly in regions such as the atmospheric boundary layer (ABL), at higher resolutions (e.g., 15 m), wind field retrieval methods based upon CDWL observations have been proposed.
Some classical wind retrieval algorithms include spectral centroid frequency estimator \cite{grey1978perceptual} and discrete spectral peak estimation \cite{hardesty1986performance}.
Nevertheless, when detecting wind fields at high altitudes (e.g., within the free atmosphere), where moving aerosol concentration is low, the signal-to-noise ratio (SNR) of the received lidar signal decreases significantly.
In low-SNR regions, traditional methods often yield retrieval results with limited accuracy, as the true Doppler frequency shift becomes obscured by noise.
Consequently, these methods often produce erroneous wind results in high-altitude atmospheric regions, resulting in a limited valid detection range.
Inspired by the success of deep learning in processing GNSS-R observables, developing data-driven approaches for retrieving wind speed over longer ranges from CDWL signals holds significant promise.

Abundant lidar observations collected by institutions worldwide support the application of AI-based methods for lidar signal processing.
CNNs have been employed for CDWL signal denoising \cite{kliebisch2022coherent, song2023denoising}, which enhances SNR of the original raw signal.
Compared to directly applying the spectral centroid estimator to the raw data, the denoised signal demonstrates an improved detection range and higher accuracy in wind speed retrieval.
Mohandes et al. \cite{mohandes2018wind} define wind retrieval as an interpolation task, utilizing deep neural network (DNN) for producing wind speeds at high altitudes based on measured values at low altitudes.
There have also been data-driven attempts for wind field prediction based on lidar observations.
Wind field prediction based on one single near-surface measurement is employed using multilayer perceptron (MLP) and other neural networks \cite{garcia2021atmospheric, garcia2023atmospheric}.
Inspired by solutions to the precipitation prediction task, Gao et al. \cite{gao2022spatio, gao2022deep} incorporates an encoder-forecaster network based on convolutional long short-term memory (LSTM) for wind field nowcasting, training and evaluating the proposed architecture on the lidar observations at Hong Kong International Airport.
In addition to wind speed detection and prediction, lidar observations can also be utilized for air pollution assessment through machine learning, as demonstrated in a recent work by Li et al. \cite{li2024integrating}.

Compared to deep learning-based ocean wind retrieval methods for GNSS-R, there has been limited research on data driven-based approaches for CDWL wind field retrieval.
Therefore, to foster the development of AI models for lidar-based wind retrieval with improved accuracy and extended range, this work focuses on state-of-the-art (SOTA) deep learning approaches for wind field retrieval leveraging CDWL observations.
We propose the \textit{first} hybrid framework designed to retrieve wind directly from raw lidar signals, establishing a benchmark for the remote sensing community.
Our model, referred to as LWFNet (Lidar-based Wind Field retrieval neural Network), integrates a modified version of vision Transformer (ViT) \cite{dosovitskiy2020image}, termed \textit{line Transformer (LiT)}, the Kolmogorov-Arnold network (KAN) \cite{liu2024kan}, and a median filter block.
The innovative structure of the line Transformer is designed with consideration of the physical characteristics of the CDWL signal power spectra.
Drawing inspirations from the Kolmogorov–Arnold representation theorem \cite{schmidt2021kolmogorov}, KAN replaces the fixed activation functions in traditional MLPs with learnable activation functions on edges, leveraging spline functions to enhance both the interpretability and accuracy of the overall network.

To evaluate the performance of LWFNet, we compare its wind retrieval results against those obtained using a traditional retrieval method, with radiosonde measurements serving as the reference standard. 
The findings reveal that LWFNet consistently outperforms the conventional method across all evaluation metrics. 
Moreover, at altitudes beyond the maximum valid retrieval range of the traditional algorithm, LWFNet continues to achieve exceptional results. 
This outcome is particularly intriguing, as LWFNet, which is trained on targets derived from the conventional retrieval method, achieves a level of accuracy that surpasses the precision of the training labels.
This phenomenon, which we termed as \textit{super-accuracy}, suggests that LWFNet may have uncovered latent patterns or relationships not fully encapsulated by the original labeling process.

% Our contributions are as follows:
Our key contributions can be summarized as:
\begin{itemize}
    \item Introduce LWFNet, the first data-driven model specifically designed for CDWL-based wind field retrieval, enhancing both the wind retrieval accuracy and detection range, thereby advancing the development of high-resolution tropospheric wind detection and monitoring.
    \item Propose a novel Transformer architecture tailored for processing CDWL power spectra, distinguishing it from the traditional Vision Transformer (ViT) structure by taking into account the physical characteristics of the spectrogram. 
    Moreover, we integrate a KAN decoder into our framework to improve both interpretability and retrieval accuracy.
    \item Conduct a series of experiments comparing the performance of the traditional method and that of LWFNet, demonstrating the latter's superior performance across all evaluation metrics.
\end{itemize}

The rest of this paper is organized as follows.
Section \ref{section:2} provides a concise overview of existing wind field retrieval approaches, encompassing both numerical and AI-based methods.
In Section \ref{section:3}, the newly proposed line Transformer and overall architecture of LWFNet shall be presented and analyzed.
Section \ref{section:4} details the data preparation process, experimental setup, evaluation metrics, and the results of both the main experiments and ablation studies.
Section \ref{section:5} offers a discussion of the results presented in Section \ref{section:4}, along with an explanation for the interesting super-accuracy phenomenon.
Finally, Section \ref{section:6} concludes the paper and provides an outlook for future research directions.

\section{Related Works}
\label{section:2}

\subsection{Numerical Wind Vector Estimation Techniques}

Generally speaking, there are two primary approaches for wind vector estimation: (1) obtain LOS wind speeds $\hat{V_{i}}$ at first, and then estimate wind vector $\hat{\boldsymbol{V}}$ through the derived results, and (2) estimate $\hat{\boldsymbol{V}}$ directly from the collected power spectra $\hat{W_{i}}$ without deriving the LOS velocity estimations \cite{smalikho2003techniques}.

The first approach encompasses several classical and widely adopted algorithms for LOS wind speed retrieval, such as spectral centroid frequency estimator \cite{grey1978perceptual}, maximum likelihood discrete spectral peak estimation \cite{levin1965power}, sine wave fitting (MSWF) and weighted sine wave fitting (WSWF) \cite{rui2019adaptive, wei2020inversion}.
On the other hand, the second approach focuses on wind velocity vector retrieval directly, including maximum likelihood for the wind vector estimation (MV ML) and maximum of the function of accumulated spectra (MFAS).
Beyond these methods, some hybrid algorithms have also been explored for wind retrieval, offering alternative approaches in this domain \cite{lin2021smoothed, zhang2024probability}.

With the advancement of more effective numerical wind vector retrieval methods, the accuracy of retrieval results has significantly improved. However, achieving precise wind retrievals in low-SNR regions remains a challenge.
Deep learning, renowned for its remarkable adaptability across various scientific domains, holds great promise in addressing this longstanding issue.

\subsection{AI-based Wind Field Retrieval Models}

Compared with numerical wind retrieval techniques, deep learning-based methods have rarely been studied.
Two most relevant works emphasize on lidar power spectrum denoising, which can also contribute to more accurate retrieval results and possibly extend the detection range.
Kliebisch et al. \cite{kliebisch2022coherent} employed SqueezeNet \cite{iandola2016squeezenet} to process lidar spectrogram, transforming denoising, which is normally a regression problem, into a classification task.
In contrast, to preserve the signal curves within the spectrogram, Song et al. \cite{song2023denoising} implemented U-Net \cite{ronneberger2015unet} for denoising. 

Experimental results from the two methods demonstrate that CNNs are efficient lidar signal denoisers, but their overall frameworks are complex and hard to train.
First, training a denoising network in a supervised manner requires preparing target data generated through lidar signal simulation algorithms \cite{henderson2005wind, abdelazim2016signal}, a process that is both time-consuming and resource-intensive.
Moreover, these simulated signals often diverge in distribution from actual measured spectrograms, as only a limited number of influencing factors are accounted for during simulation.
Additionally, after the denoising process, conventional wind retrieval algorithms must still be applied, adding another layer of complexity to the overall workflow.
Given the aforementioned limitations of the denoising task, the development of an end-to-end retrieval pipeline model based on lidar spectrograms becomes a pressing and critical challenge.

\begin{figure*}[htbp]
    \centering
    \includegraphics[width=\linewidth]{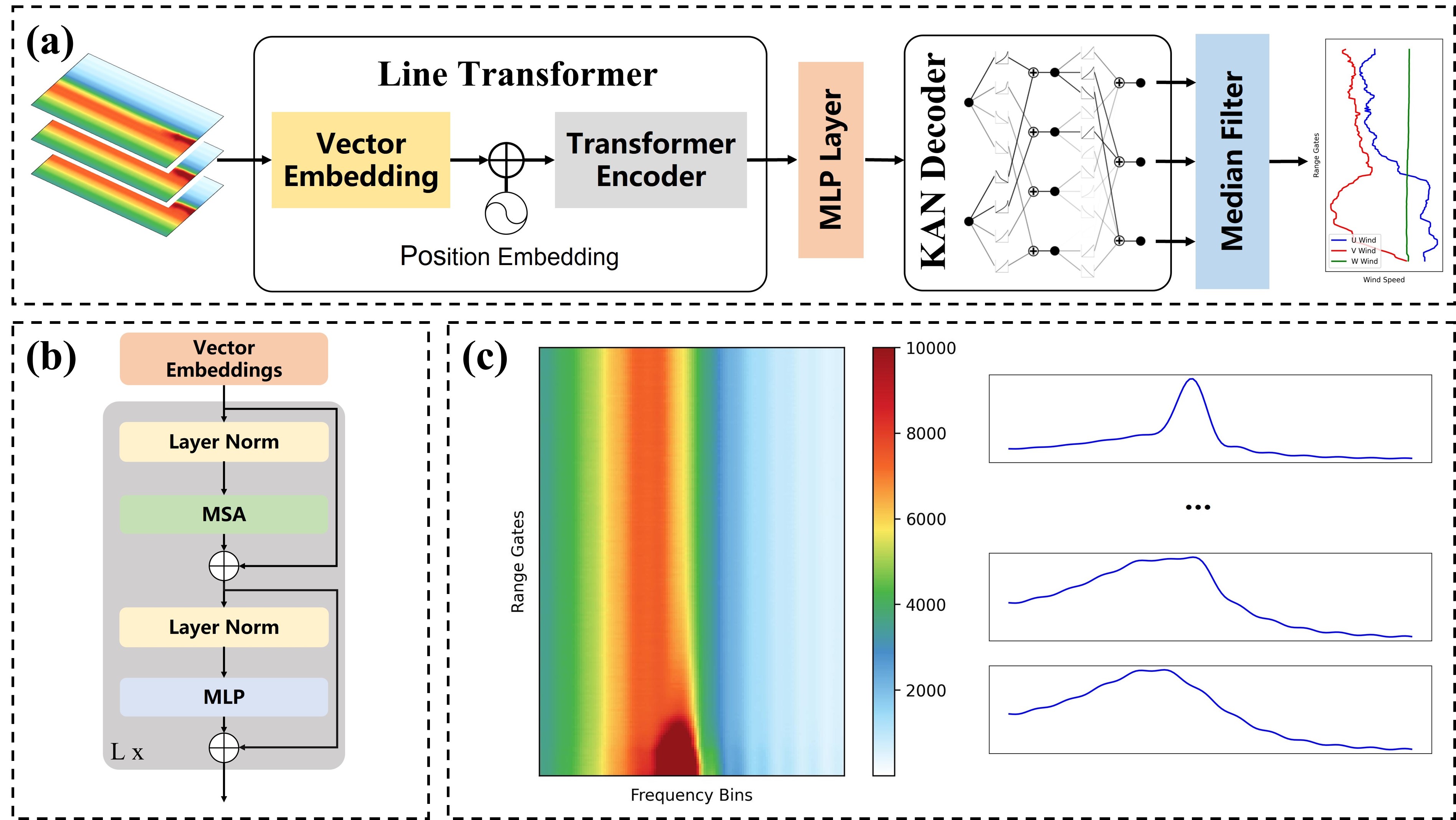}
    \caption{LWFNet framework overview and spectrogram analysis.
    (a) illustrates the backbone structure used for LWFNet signal processing, here median filter serves as an optional block, the use of which will be discussed later.
    (b) presents a detailed view of the Transformer encoder architecture.
    Finally, (c) shows the vector embedding splitting strategy, emphasizing the relative independence of backscattering signals across different range gates and showcasing the effectiveness of the unique vector embedding layer.
    }
    \label{fig:overall_framework}
\end{figure*}

On the other hand, GNSS-R-based ocean wind retrieval has witnessed the development of various AI-based techniques, such as CyGNSSnet \cite{asgarimehr2022gnss}, DDM-Former \cite{zhao2023ddm}, TCNet \cite{shi2024tcnet} and RFCN \cite{du2024deep}, which have significantly advanced ocean surface wind retrieval accuracy.
Therefore, there is an urgent need to advance the integration of deep learning with lidar signals.

We sincerely hope that LWFNet, as proposed in our work, will serve as a benchmark model for future research and inspire greater interest and engagement from researchers in the field.

\section{Methodology}
\label{section:3}

\subsection{Line Transformer}

Spectrograms received by lidar are in the shape of two-dimensional images or matrices. 
Different from natural images, however, each row of the matrix (i.e., row vector) corresponds to the power spectrum of a range gate, which is produced by applying FFT to the analog-digital converter (ADC) output.
The backscattering signal power spectrum of each range gate encapsulates information pertaining to the Doppler frequency shift, thus it can be further analyzed to obtain LOS wind velocity.

To accurately capture the intrinsic characteristics of wind speed, an elegant approach involves preserving the completeness of the power spectrum of each range gate.
CNNs, designed to extract the contextual and detailed patterns of input images, might ruin the completeness of each power spectrum.
Moreover, CNNs lack an inherent awareness of spatial locality. 
Nevertheless, in the context of backscattering spectrograms, locality of the input signal conveys information regarding detection range and frequency bin index, both of which are essential for wind speed retrieval.
In addition to CNN, ViT \cite{vaswani2017attention, dosovitskiy2020image} represents another structure adept at image processing, which aggregates global features using the Transformer encoder block.
Traditional ViTs employ a patch embedding layer, which transforms the input image into a series of tokens by dividing it into fixed-size patches.
Normally, both the height and width of patches are set as $p$.
Similar to CNNs, this embedding method might disrupt the completeness of the power spectrum.

To solve this issue, we introduce \textit{line Transformer (LiT)}, integrating the global information aggregation capability of ViT while preserving the power spectrum completeness for each range gate. 
Within LiT, we utilize a \textit{vector embedding layer} that reshapes the input spectrograms $X \in \mathbb{R}^{C \times H \times W}$ into $H$ vectors $\left \{ {\boldsymbol{x_{W}^{i}} \in \mathbb{R}^{C \cdot W} \mid i=1,...,H} \right \} $ and transforms them into $Z_{0} \in \mathbb{R}^{H \times D}$ via a convolutional layer with kernel size and stride of $(1, W)$.
Here, $C$ denotes the number of input spectrograms for wind field detection, $D$ represents the Transformer dimension, and $H$ and $W$ correspond to the number of range gates and frequency bins, respectively.
We refer to the output of this projection as \textit{vector embeddings}.
To retain positional information, position embeddings are also added to the vector embeddings.
\begin{equation}
\boldsymbol{x_{D}^{i}}=\boldsymbol{x_{W}^{i}}\boldsymbol{E}, \quad \boldsymbol{E} \in \mathbb{R}^{C \cdot W\times D}, i=1,2,...,H,
\end{equation}
\begin{equation}
Z_{0}=[\boldsymbol{x_{D}^{1};x_{D}^{2};...;x_{D}^{H} }] + PE, \quad PE \in \mathbb{R}^{H\times D},
\end{equation}
in which $\boldsymbol{x_{W}^{i}}$ and $\boldsymbol{x_{D}^{i}}$ represents the power spectrum and token corresponding to the $i$-th range gate, respectively and $PE$ denote the one-dimensional sinusoidal positional embedding \cite{vaswani2017attention}.

The vector embeddings are then sequentially processed through $L$ Transformer encoders (see Figure \ref{fig:overall_framework} (b)).
Each encoder contains Multiheaded Self-Attention (MSA) \cite{dosovitskiy2020image} and  MLP blocks.
Layernorm (LN) \cite{lei2016layer} is applied before every block, while residual connections are incorporated after each block.
Thus, the encoding process can be formulated as:
\begin{equation}
Z_{l}^{'} = \text{MSA}(\text{LN}(Z_{l-1}))+Z_{l-1}, \quad l=1,2,...,L.
\end{equation}
\begin{equation}
Z_{l} = \text{MLP}(\text{LN}(Z_{l}^{'}))+Z_{l}^{'}, \quad l=1,2,...,L.
\end{equation}

\subsection{KAN Decoder}

Currently, KAN has demonstrated its utility in a wide range of applications, including time series forecasting \cite{han2024kan4tsf} and image segmentation \cite{li2024ukan}.
In the case of LWFNet, we also integrate KAN as the decoder within the framework, replacing the traditional MLP due to KAN's superior performance and enhanced interpretability.
Generally speaking, MLP comprising of $M$ layers can be expressed as:
\begin{equation}
\text{MLP}(\boldsymbol{x})=(\boldsymbol{W_{M-1}} \circ \sigma \circ \boldsymbol{W_{M-2}} \circ \sigma \circ ... \circ \boldsymbol{W_{1}} \circ \sigma \circ \boldsymbol{W_{0}})\boldsymbol{x},
\label{eq:5}
\end{equation}
in which $\boldsymbol{W}$ and $\sigma$ denote linear and nonlinear transformations respectively.
The MLP network learns the functional mapping between input and output through a sequence of nonlinear activation functions.
Despite the expressive power assured by the universal approximation theorem \cite{vaswani2017attention}, the interior opacity within the MLP structure significantly hinders the model's interpretability.

Similar to MLP, KAN also employs fully connected structures. 
However, its functional mapping is derived through trainable one-dimensional functions, which are parametrized as splines.
Specifically speaking, the $m$-th KAN layer with $n_{m}$-dimensional input and $n_{m+1}$-dimensional output comprises of $n_{m}n_{m+1}$ learnable activation functions $\phi _{m,j,k}, j=1,...,n_{m}, k=1,...,n_{m+1}$. 
In the matrix form, we can also define a KAN layer as:
\begin{equation}
\boldsymbol{\Phi _{m} }=\begin{pmatrix} 
  \phi _{m,1,1}(\cdot)  & \phi _{m,1,2}(\cdot) & ... & \phi _{m,1,n_{m} }(\cdot)\\
  \phi _{m,2,1}(\cdot) & \phi _{m,2,2}(\cdot) & ... & \phi _{m,2,n_{m}}(\cdot)\\
  ... & ... &  & ...\\
  \phi _{m, n_{m+1} ,1}(\cdot) & \phi _{m, n_{m+1} ,2}(\cdot) & ... & \phi _{m, n_{m+1} ,n_{m}}(\cdot)
\end{pmatrix}.
\end{equation}
Therefore, when stacking all $M$ KAN layers sequentially, it can be characterized as:
\begin{equation}
\text{KAN}(\boldsymbol{x})=(\boldsymbol{\Phi _{M-1}} \circ \boldsymbol{\Phi_{M-2}} \circ ... \circ \boldsymbol{\Phi_{1}} \circ  \boldsymbol{\Phi_{0}})\boldsymbol{x}.
\label{eq:7}
\end{equation}

Thanks to the Kolmogorov-Arnold representation theorem, addition is the only multivariate function within KAN, thus the design of activation function $\phi(x)$ is crucial.
We incorporate a basis function $b(x)$ and spline function $spline(x)$ for the activation function:
\begin{equation}
\phi (x)=w_{b} b(x)+w_{s}spline(x),
\end{equation}
in which we set $b(x)=silu(x)=\frac{x}{1+e^{-x} }$ and $spline(x)= {\textstyle \sum_{i}^{}c_{i} B\text{-}sphine_{i}(x)}$.

According to the experimental results presented in \cite{liu2024kan}, the designed KAN decoder not only learns features but also optimizes them with enhanced accuracy, which is crucial for the wind field retrieval task, especially in low-SNR regions.

\subsection{LWFNet}

The overall framework of LWFNet consists of a Line Transformer for feature extraction, an MLP layer to unpatchify the output, a KAN decoder for wind field retrieval and a median filter for denoising (see Figure \ref{fig:overall_framework} (a)). This process can be summarized as:
\begin{equation}
Z_{0} = \text{VectorEmbed}(X) + PE, 
\end{equation}
\begin{equation}
Z_{l} = \text{TransformerEncoder}(Z_{l-1}), \quad l=1,\dots,L, 
\end{equation}
\begin{equation}
Z_{L}^{'} = \text{MLP}(Z_{L}), \quad Z_{L}^{'} \in \mathbb{R}^{3 \cdot H \times D}, 
\end{equation}
\begin{equation}
U_{0}, V_{0}, W_{0} = \text{KANDecoder}(Z_{L}^{'}), \quad U_{0},V_{0},W_{0} \in \mathbb{R}^{H},
\end{equation}
\begin{equation}
U, V, W = \text{MedianFilter}(U_{0}, V_{0}, W_{0}), U,V,W \in \mathbb{R}^{H},
\end{equation}

Since the KAN decoder operates exclusively on one-dimensional vectors, $Z_{L}^{'}$ is first reshaped into a sequence of $D$-dimensional vectors, with each vector corresponding to a specific range gate and directional wind component. 
After being processed by the KAN decoder, each $D$-dimensional vector generates the wind component for its associated range gate. 
As a result, we yield the U, V, and W-components of the wind field at different range gates.
Compared to the straightforward method of flattening $Z_{L}^{'}$ into a one-dimensional tensor, our approach achieves superior parameter efficiency.

Given the high spatial resolution of CDWL and the continuity of real-world wind fields, median filtering is further applied to reduce high-frequency noise in the retrieved wind speeds.

\section{Experiments and Analysis}
\label{section:4}

\subsection{Data Preparation}

\begin{figure}[htbp]
    \centering
    \includegraphics[width=\linewidth]{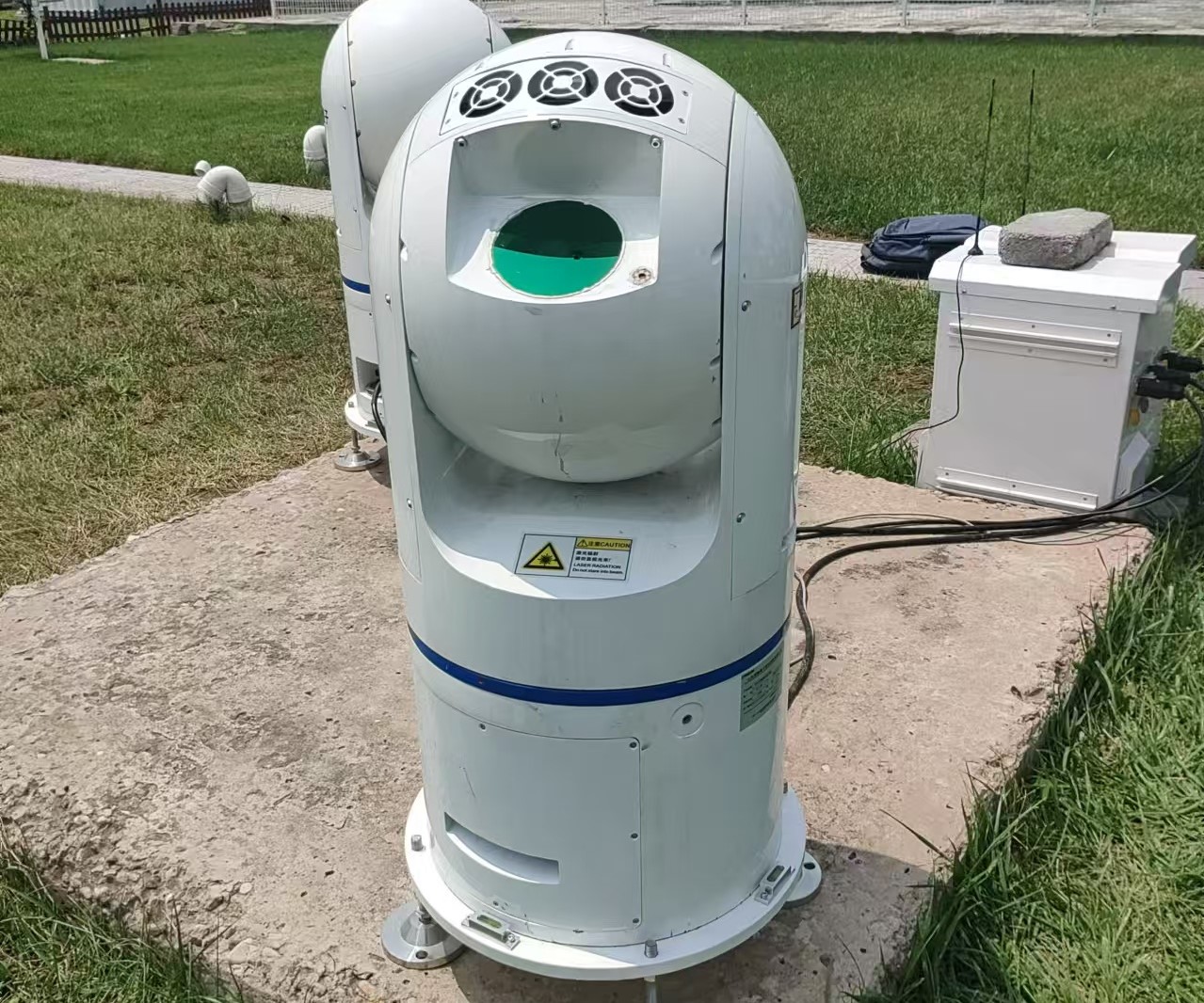}
    \caption{The ground-based coherent Doppler lidar employed for data collection and analysis.}
    \label{fig:cdwl}
\end{figure}

A 1550 nm all-fiber ground-based coherent Doppler wind lidar (see Figure \ref{fig:cdwl}) was deployed at the China Meteorological Administration (CMA) Key Laboratory of Atmospheric Sounding (KLAS, 39.83$^\circ$ N, 116.47$^\circ$ E) to collect lidar observations from December 2023 to March 2024. 
Throughout the signal acquisition process, the coherent Doppler lidar performs a velocity azimuth display (VAD) \cite{lhermitte1961precipitation} scanning technique to capture the upper atmospheric wind field.
LWFNet is designed to directly retrieve wind fields (i.e., U, V, and W components). Therefore, lidar power spectra from different scanning directions must all be considered and used as input collectively.

Since radiosonde measurements, which are often regarded as the ground truth for wind speed, are limited (i.e., typically only twice per day) and hard to obtain, we utilize wind components derived from the robust and easy-to-implement spectral centroid estimator as labels for supervised learning during the training process.
However, not all wind components derived from the spectral centroid across different range gates are reliable. 
Therefore, low-credibility wind speeds should be manually filtered out and set to NaN (Not a Number).

The filtering process considers the derived LOS wind speed, SNR, and spectrum width. Wind field results for a particular range gate are deemed valid only if power spectra from all scanning directions at that range gate satisfy the following conditions:
(1) The derived LOS wind speed is less than 42 $m/s$;
(2) The SNR is greater than -35 $dB$;
(3) The spectrum bandwidth is between 0.5 and 7.5 (in units of $\Delta f$, the frequency resolution of frequency bins);
(4) If condition in (3) is not met, but the SNR exceeds -25 $dB$, the results are still considered credible.
For convenience, we refer to the range gates meeting these criteria as \textit{high-SNR regions} and those that do not as \textit{low-SNR regions}, as SNR is typically the most influential factor in the filtering process.

\renewcommand{\arraystretch}{1.5}
\begin{table*}[!ht]
\caption{Average Evaluation Metrics for Spectral Centroid and LWFNet across 32 test instances. Pearson CC denotes the Pearson Correlation Coefficient. 
}
\label{metrics_comp_chart}
\centering
\begin{tabular}{cc|ccc|ccc}
\toprule
\multirow{2}{*}{Method} & \multirow{2}{*}{Region of Interest} & \multicolumn{3}{c|}{Horizontal Wind Speed (m/s)} & \multicolumn{3}{c}{Horizontal Wind Direction ($^\circ$)} \\
\cline{3-5} 
\cline{6-8}
       &    & RMSE ($\downarrow$) & MAE ($\downarrow$) & Pearson CC ($\uparrow$) & RMSE ($\downarrow$) & MAE ($\downarrow$) & Pearson CC ($\uparrow$)\\
\midrule
Spectral Centroid Estimator & \multirow{2}{*}{\parbox{2.8cm}{\centering High-SNR Region}}   & 0.885  & 0.697   & 0.899   & 16.899    & 9.931   & 0.797   \\
% \hline
% LWFNet &  & 0.862  &   0.684   &  0.905   &   12.559   &   8.351   &   0.856 \\
LWFNet  &  & 0.795  &   0.634   &  0.920   &   11.167   &   7.576   &   0.877 \\
\hline
% LWFNet & \multirow{2}{*}{\parbox{2.8cm}{\centering Entire Region}} & 1.785  &   1.250   &  0.785   &   20.986   &   13.573   &   0.787 \\
LWFNet  &  Entire Region  & 1.425  &  1.052   &  0.845   &   15.181   &   10.236   &  0.859 \\
\bottomrule
\end{tabular}
\end{table*}

\begin{figure*}[htbp]
    \centering
    \includegraphics[width=\linewidth]{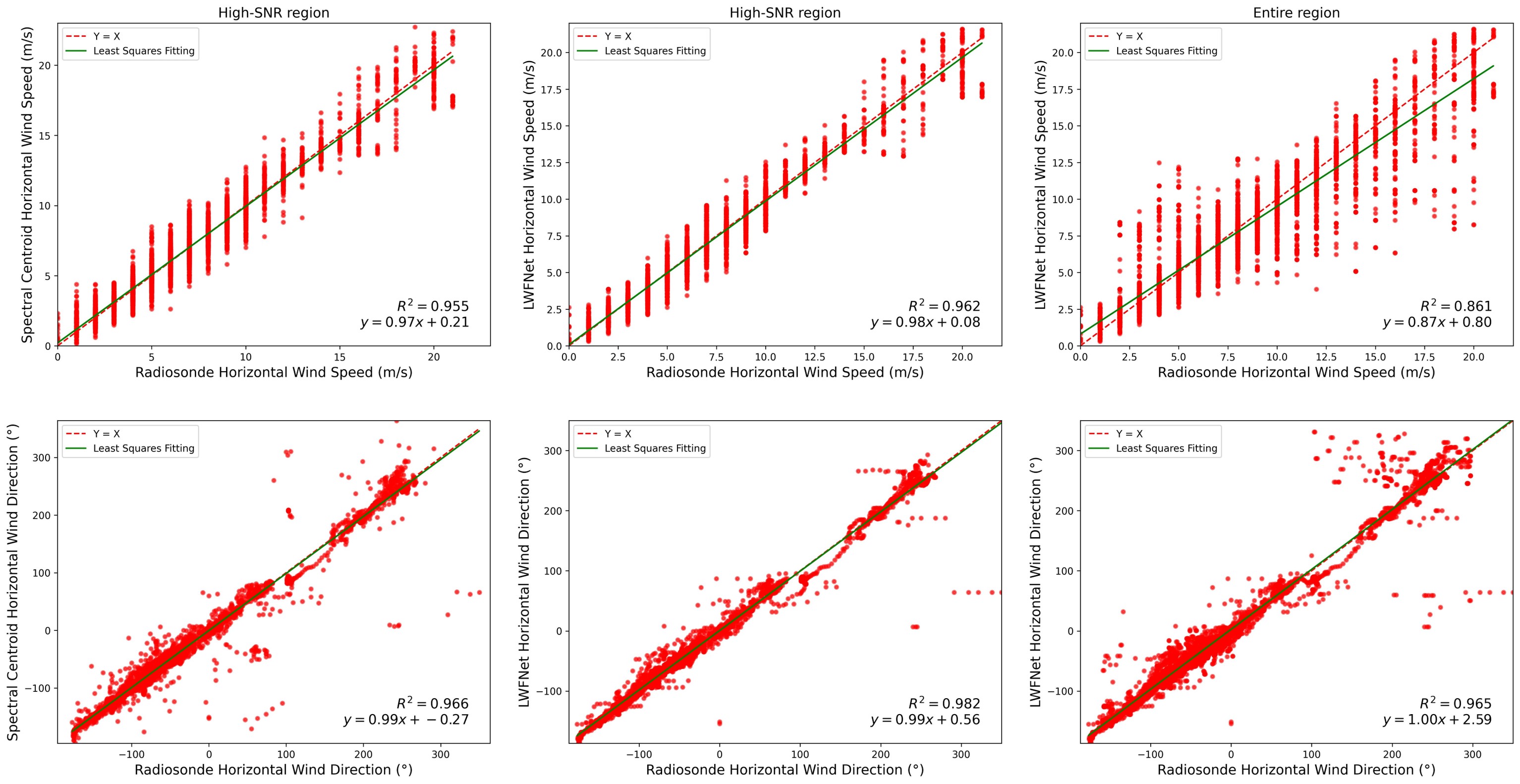}
    \caption{
    Scatter plots for the spectral centroid estimator and LWFNet, compared against radiosonde measurements, are presented for all 32 test instances in March 2024.
    The subtitles at the top of the columns indicate the specific region of interest tested.
    The first and second rows display the scatter plots for horizontal wind speed and horizontal wind direction, respectively.
    The $R^{2} $ values and least squares fitting line formula are illustrated at the bottom right corner of each plot.
    Note: (1) The speed resolution of radiosonde measurement data is 1 $m/s$, thus the scatter plot for horizontal wind speed takes the form of discrete columns.
    (2) Azimuth angles range from 0 $^\circ$ to 360$^\circ$ within a closed loop. To address situations where small differences in wind direction correspond to large absolute angular differences (e.g., 1 $^\circ$ and 359 $^\circ$), here we expand the conventional wind direction range from 0$^\circ$ to 360$^\circ$ to the range from -180$^\circ$ to 360$^\circ$.}
    \label{fig:6_density_plot}
\end{figure*}

\subsection{Implementation Details}

For the lidar used in our experiment, each spectrogram consists of 220 range gates and 128 frequency bins, and the system employs a conventional 3-directional VAD scan.
Therefore, the input tensors for LWFNet have the shape of $C=3$, $H=220$ and $W=128$.
Consequently, the labels for each wind component also have a length of 220. 
We set the Transformer dimension to $D=128$ and use $L=4$ Transformer encoder blocks for the Line Transformer.
The decoder consists of a 2-layer KAN structure, with the number of neurons in the middle layer set to 32.
The median filtering block used for signal postprocessing applies a spatial window with a length of 7 range gates.
The range gate spatial resolution is 15 $m$, but considering the mirror effect at low altitudes and fiber optic circuit losses within the lidar, the results from 20 to 40 range gates are invalid. Therefore, the theoretical maximum wind field detection range is 2700 to 3000 meters above the ground.
While the maximum wind detection range of conventional wind retrieval methods often falls short of the theoretical limit, LWFNet is designed to extend this range

For the training of LWFNet, we use lidar observations collected from December 2023 to February 2024 as training and validation sets, while observations from March 2024, when radiosonde data is available, are used as the test set.
In total, we obtain 32 instances of radiosonde ground truth observations, each containing information on horizontal wind speed and direction.
Therefore, in evaluating and comparing the performance of LWFNet and the spectral centroid estimator outputs, we primarily focus on the horizontal wind components.
Here, we define due north as a 0-degree horizontal wind direction and measure angles clockwise.

During the model training process, We use mean absolute error (MAE) as the loss function.
Targets containing NaN values are excluded from the loss function computation, meaning a mask is applied to these NaN values to ensure they do not contribute to the loss calculation.
AdamW optimizer \cite{loshchilov2017decoupled} with an initial learning rate of $1.0 \times 10^{-3}$ is utilized.
Additionally, a cosine annealing schedule is applied, featuring a single cosine cycle over 100 epochs with a minimum learning rate of $1.0 \times 10^{-4}$.
LWFNet is trained for a total of 1000 epochs with a batch size of 1024, and the best model is selected prior to overfitting.

\begin{figure*}[htbp]
    \centering
    \includegraphics[width=\linewidth]{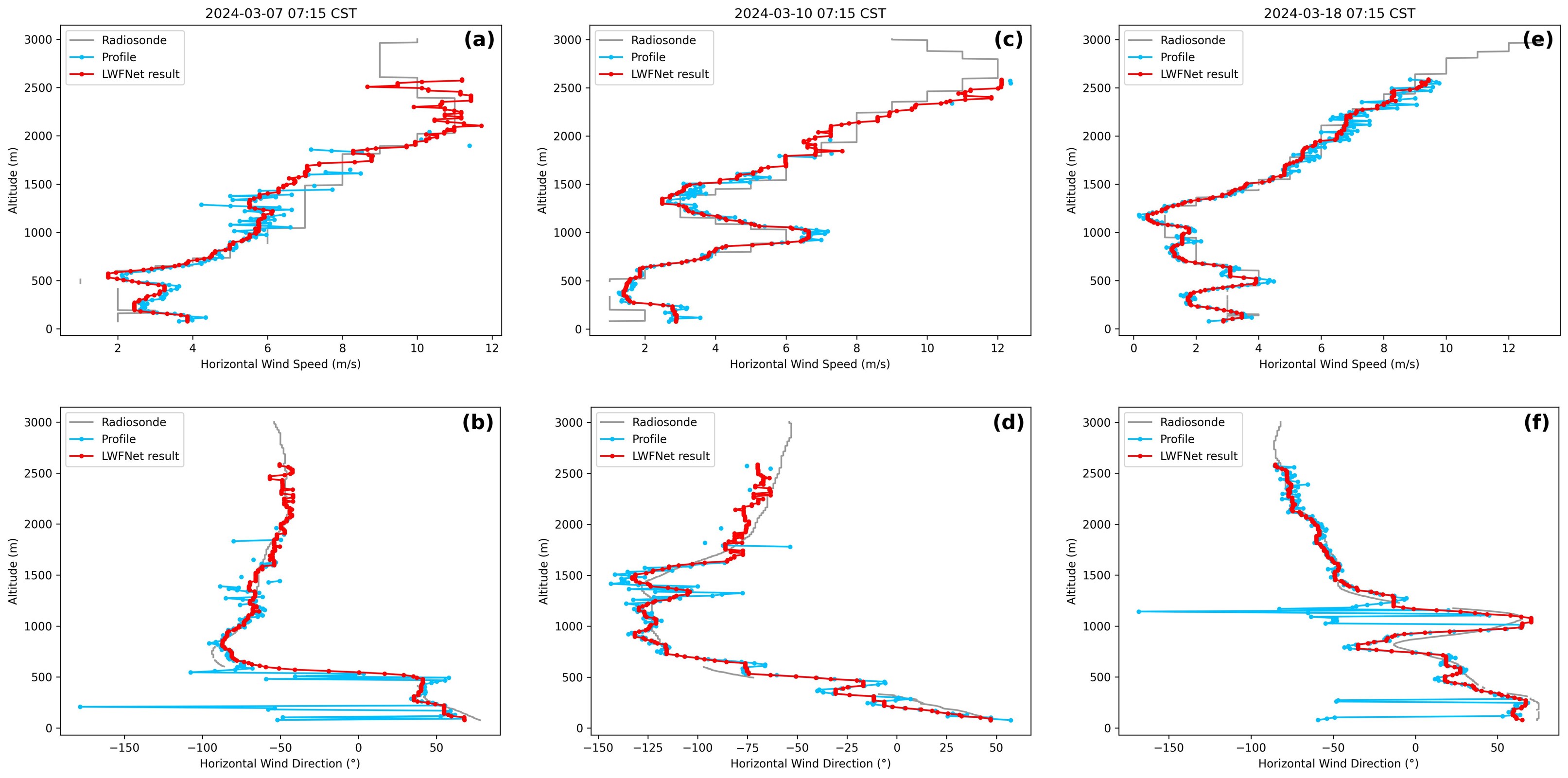}
    \caption{Wind field retrieval results of LWFNet and the spectral centroid estimator (denoted as "Profile"), compared with radiosonde ground truth. 
    The subtitles at the top of the columns indicate the testing dates.
    The first and second rows present the evaluation results for horizontal wind speed and horizontal wind direction, respectively.
    Note: CST refers to China Standard Time and will be used as such throughout the text.}
    \label{fig:6_inst}
\end{figure*}

\subsection{Evaluation Metrics}

Following the convention used by meteorologists, we first transform the U and V-components of the wind field into horizontal wind speed and direction before calculating the evaluation metrics.
Using the spectral centroid as our baseline, we evaluate the retrieval results of LWFNet from two perspectives: (1) the performance in regions where the spectral centroid algorithm produces valid retrievals, and (2) the overall performance across all range gates.

In both cases, the performance of LWFNet is quantified using RMSE (Root Mean Square Error), MAE (Mean Absolute Error), and Pearson correlation coefficient.
Let $v$, $\hat{v}$, and $N$ represent the ground truth and retrieved wind speed or direction, and the number of range gates taken into account, respectively.
The evaluation metrics can be defined as follows:
\begin{equation}
\text{RMSE}(v,\hat{v})=\sqrt{\frac{1}{N} \sum_{i=1}^{N} (v-\hat{v})^{2} } ,
\end{equation}
\begin{equation}
\text{MAE}(v,\hat{v})=\frac{1}{N}{\sum_{i=1}^{N} \left | v-\hat{v}  \right |} ,
\end{equation}
\begin{equation}
\text{r}(v,\hat{v})=\frac{ {\textstyle \sum_{i=1}^{n}} (v_{i}-\bar{v})(\hat{v}_{i}-\bar{\hat{v}})}{\sqrt{ {\textstyle \sum_{i=1}^{n}(v_{i}-\hat{v}_{i})^2 {\textstyle \sum_{i=1}^{n}} (\hat{v}_{i}-\bar{\hat{v}})^2} } } ,
\end{equation}
in which $\bar{v}$ and $\bar{\hat{v}}$ denote the average radiosonde measurements and average retrieved wind speed or direction, respectively.

\begin{figure*}[htbp]
    \centering
    \includegraphics[width=\linewidth]{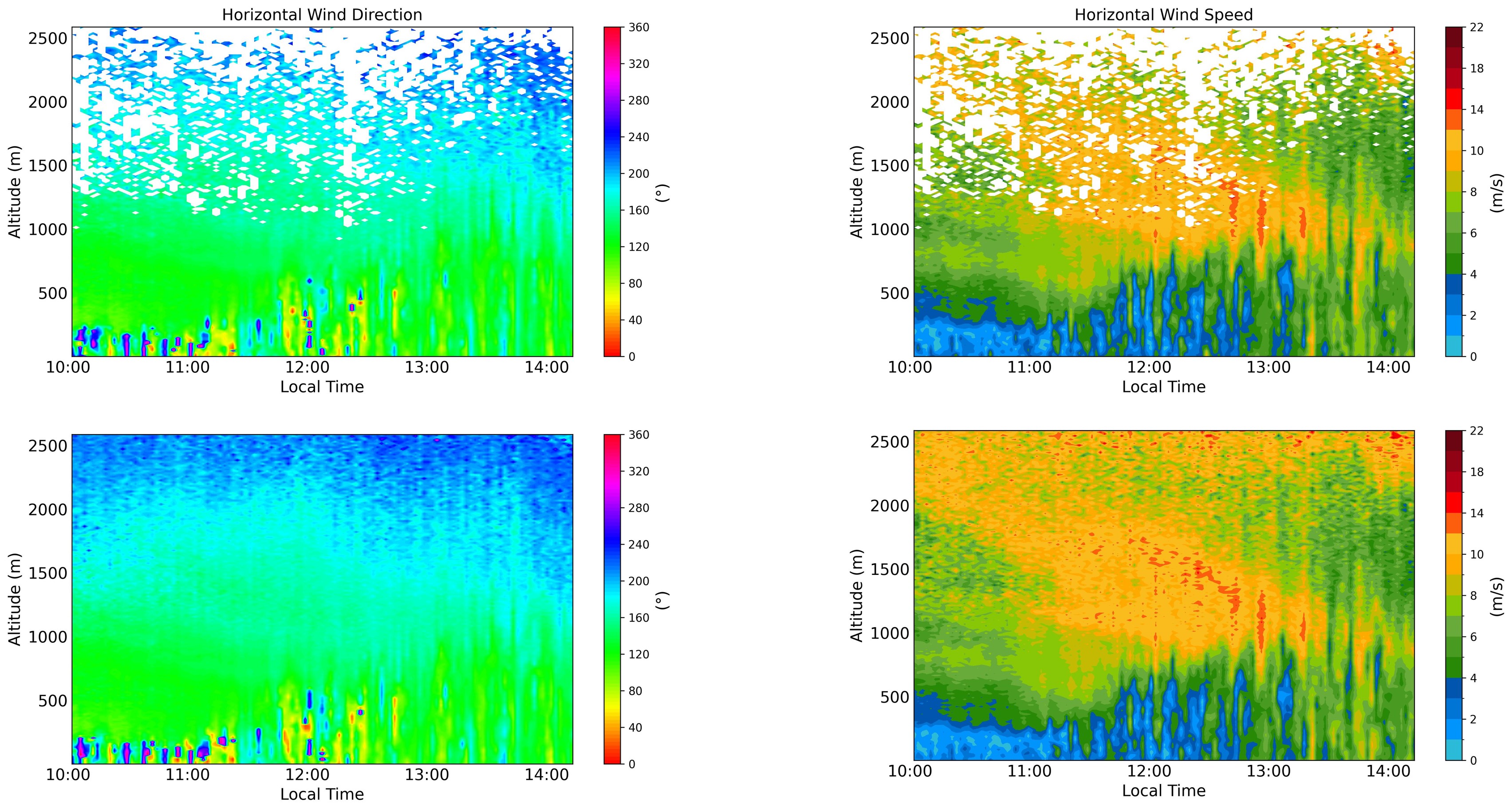}
    \caption{Wind field retrieval results of LWFNet and the spectral centroid estimator on March 14th, 2024, from 10:00 to 14:00 CST, are presented.
    The first and second rows display the retrieval results from the spectral centroid algorithm and LWFNet, respectively.
    }
    \label{fig:4_continuous}
\end{figure*}

\renewcommand{\arraystretch}{1.5}
\begin{table*}[!ht]
\caption{Average Evaluation Metrics for ResNet, ViT and LWFNet across 32 test instances.
Pearson CC denotes the Pearson Correlation Coefficient. 
Note: The \textbf{best} performance metrics are bolded.
}
\label{sota_comp_chart}
\centering
\begin{tabular}{cc|ccc|ccc}
\toprule
\multirow{2}{*}{Model} & \multirow{2}{*}{Region of Interest} & \multicolumn{3}{c|}{Horizontal Wind Speed (m/s)} & \multicolumn{3}{c}{Horizontal Wind Direction ($^\circ$)} \\
\cline{3-5} 
\cline{6-8}
       &    & RMSE ($\downarrow$) & MAE ($\downarrow$) & Pearson CC ($\uparrow$) & RMSE ($\downarrow$) & MAE ($\downarrow$) & Pearson CC ($\uparrow$)\\
\midrule
ResNet & \multirow{3}{*}{\parbox{2.8cm}{\centering High-SNR Region}}  & 1.435  & 1.167   & 0.858   & 14.259    & 10.188   & 0.811   \\
ViT &    & 0.910  & 0.698   & 0.915   & 14.686    & 8.933   & 0.771   \\
\textbf{LWFNet (Ours)}  &  & \textbf{0.795}  &   \textbf{0.634}   &  \textbf{0.920}   &   \textbf{11.167}   &   \textbf{7.576}   &   \textbf{0.877} \\
\hline
ResNet & \multirow{3}{*}{\parbox{2.8cm}{\centering Entire Region}}  & 2.202  & 1.718   & 0.791   & 19.032   & 14.191   & 0.766  \\
ViT &  & 1.551  &   1.172   &  \textbf{0.861}   &   18.565   &   12.752   &   0.843 \\
\textbf{LWFNet (Ours)}  &   & \textbf{1.425}  &  \textbf{1.052}   &  0.845  &   \textbf{15.181}   &   \textbf{10.236}   &  \textbf{0.859} \\
\bottomrule
\end{tabular}
\end{table*}

\subsection{Experimental Results}

To evaluate the effectiveness of our proposed LWFNet, we compare the results of the spectral centroid estimator and LWFNet across 32 test instances. 
For each test instance, due to the inherent bias between the coordinate systems of the ground-based lidar and radiosonde, we introduce a fixed bias to both the spectral centroid and LWFNet retrieval results. 
This bias is determined by minimizing the mean squared error between the wind retrieval sequence and the corresponding radiosonde measurement sequence in high-SNR regions.
All subsequent results presented account for this bias adjustment.

Compared to the radiosonde ground truth observations, Table \ref{metrics_comp_chart} reports the average evaluation metrics for all test samples.
Experimental results indicate that the retrieval outputs from the spectral centroid estimator in low-SNR regions exhibit significant fluctuations resembling high-frequency noise, therefore only the spectral centroid results within high-SNR regions are presented.
As shown in the table, within high-SNR regions, LWFNet surpasses the spectral centroid estimator on all metrics.
This is a particularly interesting result, as our network is trained solely on spectral centroid retrieval results, and the fixed bias is determined based on the distributional difference between spectral centroid outputs and radiosonde observations.
Notably, neither of these processes incorporates any additional information sources to enhance the accuracy of the trained model, unlike \cite{kliebisch2022coherent} and \cite{song2023denoising}, which augmented the training dataset with simulated spectrograms, thereby introducing extra labeled data.
To the best of our knowledge, this is the \textit{first} time that such a \textit{super-accurate} phenomenon is discovered within the deep learning domain.

Across the entire wind field detection range (i.e., 220 range gates altogether), LWFNet continues to demonstrate exceptional performance, with errors falling within the acceptable range for meteorological standards.
The RMSE and Pearson correlation coefficient of LWFNet across the entire wind field detection range even surpass those of the spectral centroid estimator in high-SNR regions.

Similar to the quantitative analysis in Table \ref{metrics_comp_chart}, Figure \ref{fig:6_density_plot} provides scatter plots for the two methods in different regions of interest.
For clarity, it is important to state that scatter plots for horizontal wind velocity take the form of discrete columns, because the speed resolution of radiosonde observation data is 1 $m/s$. 
The $R^{2}$ value and least mean squares fitting line formula are presented at the bottom right corner of each plot.
Consistent with the findings in Table \ref{metrics_comp_chart}, we observe that LWFNet outperforms the spectral centroid estimator in high-SNR regions, delivering more accurate results.
Here we examine the performance of LWFNet in low-SNR regions more closely.
We find that horizontal wind direction is less affected by SNR compared to wind velocity, exhibiting a smaller increase in error variance compared to high-SNR regions.
This is another intriguing conclusion, but the underlying mechanism remains unclear at present.

To better study and analyze retrieval results, Figure \ref{fig:6_inst} visualizes three test cases: March 7th at 7:15, March 10th at 7:15, and March 18th at 7:15, respectively.
Setting radiosonde observations as ground truth, the results of the spectral centroid estimator and LWFNet are illustrated.
The first two columns of Figure \ref{fig:6_inst} are two examples of LWFNet extending the credible wind field detection range.
High-velocity winds, especially at high altitudes, are inherently difficult to detect accurately due to the skewed distribution of training data, where low-velocity wind fields overwhelmingly dominate the samples.
However, LWFNet is capable of handling this task with remarkable ease.
Moreover, the first and last columns of Figure \ref{fig:6_inst} also display examples of accurate ground-level wind field detection.
Due to the presence of severe near-ground wind shear, ground-level wind velocity detection remains a significant challenge for CDWL.
Again, LWFNet addresses this problem with enhanced accuracy.

To display the temporal characteristics of LWFNet, Figure \ref{fig:4_continuous} provides retrieval results of both spectral centroid and LWFNet for horizontal wind direction and speed, recorded on March 14th, 2024, between 10:00 and 14:00 CST.
As plotted in the figure, LWFNet seamlessly fills the missing data, avoiding abrupt changes in the wind field within the spatial and temporal domain.
This further demonstrates its capability in long-period wind retrieval.

\renewcommand{\arraystretch}{1.5}
\begin{table*}[!ht]
\caption{Architectural details of model variants used in the ablation study
}
\label{ablation_transformer_structure}
\centering
\begin{tabular}{c|ccc|cccc}
\toprule
\multirow{2}{*}{Model} & \multicolumn{3}{c|}{Model Components} & \multicolumn{4}{c}{Structure Parameters} \\
\cline{2-4} 
\cline{5-8}
    & Transformer Type & Decoder Type & Median Filter & Patch Size & Sequence Length & Transformer Dimension & KAN Dimension 
    % $[n_{0},..., n_{M-1}]$ 
    \\
\midrule
ViT-KAN  & ViT & KAN & \ding{51} & (10, 8)  & 352 & 128 & [220, 32, 1] \\
LiT & LiT & MLP & \ding{51} &  (1, 128)  & 220 & 128 & - \\
LiT-KAN & LiT & KAN & \ding{55} &  (1, 128)  & 220 & 128 & [220, 32, 1] \\
LWFNet  & LiT & KAN & \ding{51} & (1, 128)  & 220 & 128 & [220, 32, 1] \\
\bottomrule
\end{tabular}
\end{table*}

\renewcommand{\arraystretch}{1.5}
\begin{table*}[!ht]
\caption{Average Evaluation Metrics for LWFNet and other model variants across 32 test instances.
Pearson CC denotes the Pearson Correlation Coefficient. 
Note: The \textbf{best} and \underline{second-best} metrics are highlighted.
}
\label{ablation_comp_chart}
\centering
\begin{tabular}{cc|ccc|ccc}
\toprule
\multirow{2}{*}{Model} & \multirow{2}{*}{Region of Interest} & \multicolumn{3}{c|}{Horizontal Wind Speed (m/s)} & \multicolumn{3}{c}{Horizontal Wind Direction ($^\circ$)} \\
\cline{3-5} 
\cline{6-8}
       &    & RMSE ($\downarrow$) & MAE ($\downarrow$) & Pearson CC ($\uparrow$) & RMSE ($\downarrow$) & MAE ($\downarrow$) & Pearson CC ($\uparrow$)\\
\midrule
ViT-KAN & \multirow{4}{*}{\parbox{2.8cm}{\centering High-SNR Region}} & 0.862  &   0.684   &  0.893   &   11.751   &   8.052   &   \underline{0.864} \\
LiT  &  & \underline{0.817}  &   \underline{0.658}   &  \underline{0.915}  &  \textbf{10.920}   & \textbf{7.462}  &  0.860 \\
LiT-KAN  &  & 0.862  &   0.684   &  0.905   &   12.559   &   8.351   &   0.856 \\
LWFNet  &  & \textbf{0.795}  &   \textbf{0.634}   &  \textbf{0.920}   &   \underline{11.167}   &   \underline{7.576}   &   \textbf{0.877} \\
\hline
ViT-KAN &  \multirow{4}{*}{\parbox{2.8cm}{\centering Entire Region}}  & 1.575  &   1.182   &  \underline{0.827}  &   16.555   &   12.073   &   \textbf{0.873} \\
LiT  &  & \underline{1.537}  &   \underline{1.136} &  \underline{0.827}  &  \textbf{15.026}  &  \underline{10.440} & \underline{0.864} \\
LiT-KAN &    & 1.785  &   1.250   &  0.785   &   20.986   &   13.573   &   0.787 \\
LWFNet  &   & \textbf{1.425}  &  \textbf{1.052}   & \textbf{0.845}   &   \underline{15.181}   &   \textbf{10.236}   &  0.859 \\
\bottomrule
\end{tabular}
\end{table*}

\subsection{Deep Learning Model Comparison}

In order to further verify the superiority of our model over other deep learning approaches, Table \ref{sota_comp_chart} presents a comparison of LWFNet's performance with SOTA models, specifically a convolution-based (i.e., ResNet) and a Transformer-based (i.e., ViT) model.
Within our experiments, ResNet follows the 18-layer architecture proposed in \cite{hekaiming2016deep}, with four building blocks serving as the backbone.
ViT adheres to the traditional paradigm, utilizing a patch embedding layer with patch size of $(10, 8)$ and an MLP as the decoder structure.
Both ResNet and ViT include a median filter for post-processing the wind retrieval components.
Following the same training and evaluation protocol used for LWFNet, we conduct identical testing experiments for these models.
The hyperparameters, such as batch size and learning rate, are kept consistent with those used during the training of LWFNet.

Comprehensively speaking, LWFNet outperforms both ResNet and ViT across nearly all metrics, highlighting its superior performance in deep learning for wind retrieval tasks. 
Moreover, we observe that the Transformer structure is more suitable for the wind retrieval task than convolution, suggesting that the lack of spatial locality in CNNs may hinder their performance in this context.
In fact, this is not a particularly surprising result, as the local connectivity of CNNs also limits their ability to generate feature maps that encapsulate the full signal information of all range gates at early stages.

\subsection{Ablation Study}

To study how each component of the LWFNet architecture contributes to the wind retrieval results, additional comparative experiments are conducted.
We compare the performance of LWFNet with three variants: ViT-KAN,  which replaces the LiT block of LWFNet with the conventional ViT; LiT, which substitutes the KAN component of LWFNet with an MLP decoder; and LiT-KAN, which removes the median filtering block from LWFNet.
The architectures of these variants are detailed in Table \ref{ablation_transformer_structure}, and the corresponding testing results are presented in Table \ref{ablation_comp_chart}.

\textbf{Effect of Line Transformer.}
Comparing the conventional ViT-based (e.g., ViT-KAN and ViT from Table \ref{sota_comp_chart}) with LiT-based models (e,g., LiT and LWFNet), we find that the latter constantly outperforms the former in both horizontal wind speed and direction retrieval, suggesting the effectiveness of the vector embedding layer.
This is a thought-provoking result, as LiT outperforms both ViT and ViT-KAN on many metrics, despite having the fewest number of Transformer embeddings and, consequently, the least number of model parameters compared to the other Transformer-based models.
The superior performance of LiT largely relies on its excellent signal processing structure, demonstrating the importance of image tokenization in lidar-based wind retrieval.

\textbf{Effect of KAN Decoder.}
The role of the KAN decoder is also obvious, owing to its strong interpretability and approximation ability.
With only a 2-layer KAN decoder utilized, the accuracy of the results, as measured by RMSE and MAE, has improved in most cases.
This conclusion is supported not only by comparing the performance of LiT and LWFNet, but also by contrasting ViT from Table \ref{sota_comp_chart} and ViT-KAN.
However, in terms of the Pearson correlation coefficient, the contribution of the KAN decoder is less pronounced, and the underlying mechanism of this behavior warrants further investigation.

\textbf{Effect of Median Filter.}
The performance comparison between LiT-KAN and LWFNet suggests that, for high spatial resolution instruments, such as coherent Doppler lidar, filtering out high-frequency components manually can lead to a significant improvement in retrieval results.
Typically, continuity characteristics of wind fields imply that abrupt shifts do not occur on meter-scale, unless influenced by human intervention.
Hence, median filtering is a very \textit{safe} operation for refining the retrieval results.

\begin{figure}[htbp]
    \centering
    \includegraphics[width=\linewidth]{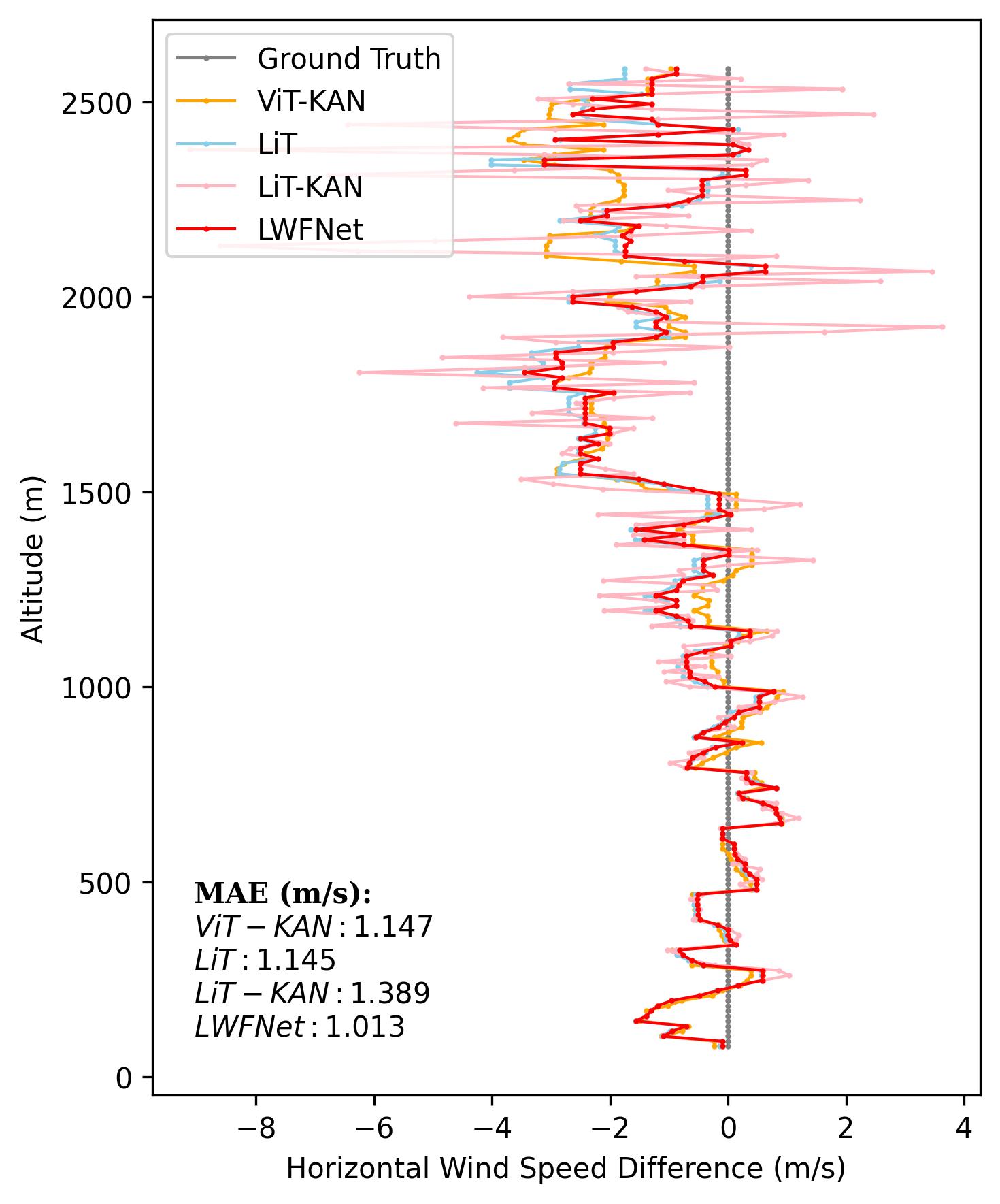}
    \caption{Wind differences between the retrieval results of multiple deep learning model variants and radiosonde ground truth at different heights, on March 14th, 2024, at 7:15 CST.
    The average MAE values for each model are presented at the bottom left of the image.
    }
    \label{fig:1_ablation_speed}
\end{figure}

To visualize the wind retrieval capabilities of different deep learning model variants, Figure \ref{fig:1_ablation_speed} illustrates the horizontal wind speed differences between the retrieval results of various models and radiosonde measurements at different heights on March 14th, at 7:15 CST.

\section{Discussion}
\label{section:5}

Although our proposed LWFNet outperforms the spectral centroid algorithm, it is worth briefly discussing some limitations of our work and possible reasons for the super-accurate phenomenon mentioned in Section \ref{section:4}.

\begin{figure}[htbp]
    \centering
    \includegraphics[width=\linewidth]{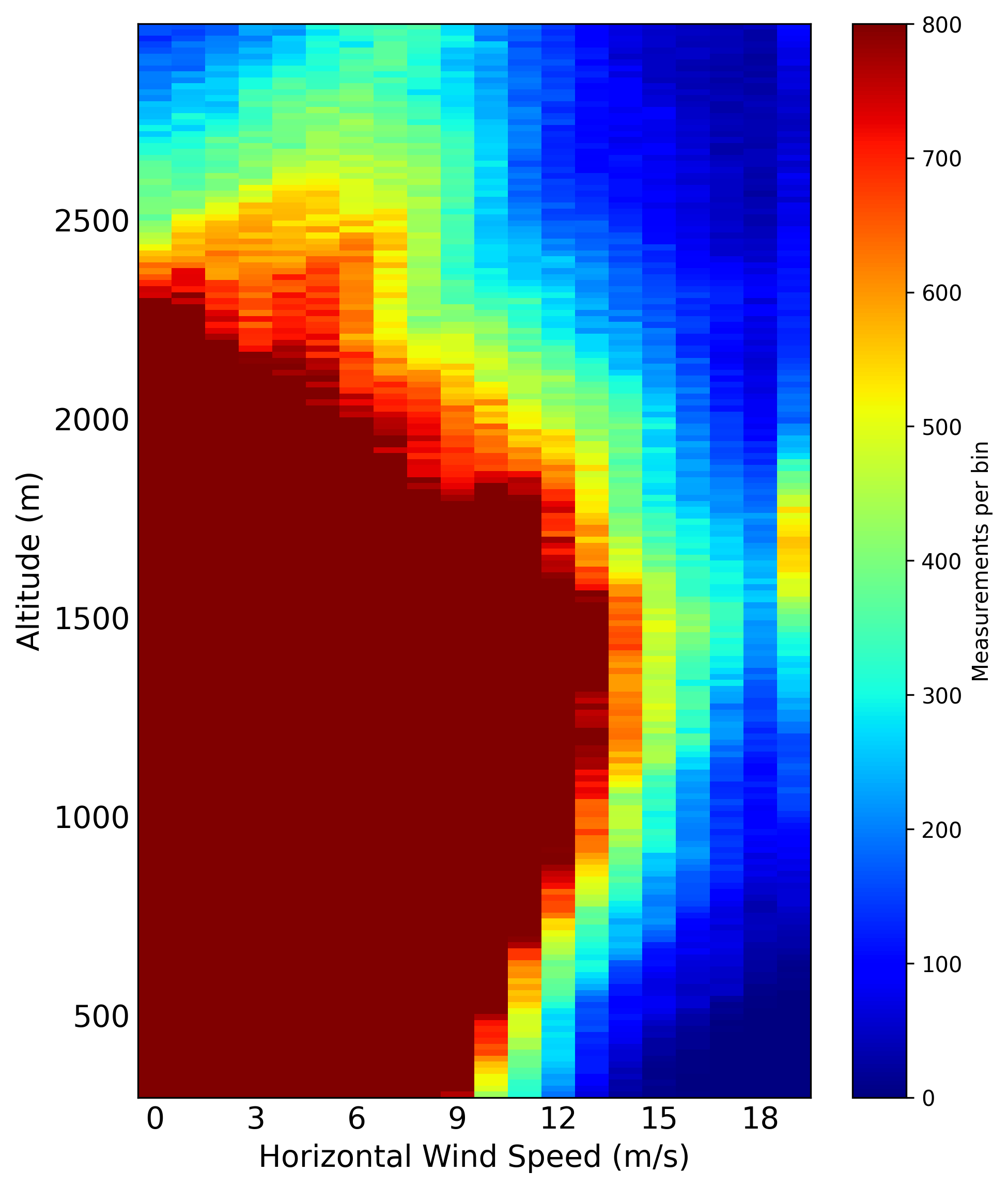}
    \caption{Distribution of labeled horizontal wind speed in the training and validation datasets.
    The color bar measures the number of samples at the corresponding altitude and wind speed.
    }
    \label{fig:discuss_distribution}
\end{figure}

To start with, we emphasize the consequences caused by the lack of enough observation data.
Since all of our experimental conclusions are based on the 32 available radiosonde measurements within the same month, which represent a limited range of wind field types, there could be potential bias in our findings.
To further assess the comprehensive performance of LWFNet, more extensive test experiments should be carried out.
Additionally, our training dataset includes no more than three months of collected power spectra, therefore seasonal or other periodic factors were not incorporated into the training process.
It would be beneficial to expand both the training and testing datasets to capture a broader range of conditions.
In Figure \ref{fig:discuss_distribution}, we illustrate the distribution of labeled horizontal wind speed in the training and validation datasets.
It is evident that low-velocity winds dominate a large proportion of the dataset, while high-altitude wind labels are significantly underrepresented.
Collecting additional training data would not only enable the model to learn from a more diverse set of wind types, but also provide more labeled instances for supervised training.

We shall also explore the underlying reasons behind the intriguing super-accurate phenomenon.
Like most numerical methods, the wind retrieval process of the spectral centroid estimator does not take the power spectrum of neighboring range gates into account.
One possible reason for super-accuracy is that LWFNet considers backscattering signals across all range gates and learns their connections and relations.
CNNs take the neighboring bins into account, but only feature maps of the last few stages would gain an overview of the entire input spectrogram.
ViT-based models, on the other hand, are excellent in dealing with natural images because they aggregate global information through the encoder blocks, but also corrupt lidar signal by reshaping it into square-shaped patches.
Hence, most deep learning models in Table \ref{sota_comp_chart} have shown superior performance than the spectral centroid algorithm by considering the entire spectrogram for wind retrieval, with the novel architecture of LWFNet emerging as the most effective solution.

Furthermore, the inherent smoothing effect of deterministic models may assist in wind retrieval.
As shown in Table \ref{ablation_comp_chart}, median filtering contributes to more accurate retrieval results by acting as a low-pass filter. 
This is because wind fields exhibit strong spatial and temporal continuity, with very few high-frequency components.
Similarly, the smoothing characteristic might also help neural networks generate low-frequency results that align with the natural behavior of real-world wind.

Finally, the training strategy is also worth discussing.
Different from numerical methods, deep learning models retrieve wind fields within low-SNR regions by accounting for high-SNR region examples, demonstrating that masking a proportion of targets enhances the decoding capability of models. 
Inspired by the MAE (masked autoencoder) structure \cite{he2022MAE}, our approach involves masking targets during training to facilitate the development of more efficient decoders.
The effectiveness of such an approach is still unclear, and further validation experiments are needed.

\section{Conclusion}
\label{section:6}

We introduce LWFNet, the first hybrid deep learning model for wind retrieval based on coherent Doppler lidar.
In lidar signal processing, our proposed line Transformer outperforms the traditional vision Transformer upon multiple metrics, demonstrating its superior effectiveness for this task.
Further, by incorporating KAN decoder within our structure, LWFNet offers more accurate and interpretable retrieval results.
Compared with the conventional wind field retrieval algorithm (i.e., spectral centroid), LWFNet outperforms it on nearly all metrics for horizontal wind speed and direction in high-SNR regions.
In addition, the results it generates across the entire region also adhere to the acceptable limits of meteorological standards.
The robustness of our model is further demonstrated by comparing its performance against other SOTA data-driven models.

During the testing stage of LWFNet, we observe the \textit{super-accurate} phenomenon, by means that the model's wind retrieval results significantly exceed the accuracy of the training targets.
To the best of our knowledge, such an occurrence has not been reported elsewhere.
We attribute this exceptional performance to the characteristics of the Transformer architecture, the inherent smoothing effect of our proposed model, and the masking strategy employed during training.

Considering the limited research done on CDWL-based deep learning models for wind retrieval, we sincerely hope that LWFNet shall serve as a benchmark model, advancing research in high-resolution wind field detection and promoting the development of AI-based methods for broader applications.

\section{Aknowledgement}
\label{section:7}

This work is supported by Shanghai Artificial Intelligence Laboratory, Shandong Guoyao Quantum Lidar Company Ltd. and Anhui Institute of Meteorological Sciences.

\IEEEpeerreviewmaketitle

\ifCLASSOPTIONcaptionsoff
 \newpage
\fi

% Generated by IEEEtranN.bst, version: 1.14 (2015/08/26)


\begin{thebibliography}{57}
\providecommand{\natexlab}[1]{#1}
\providecommand{\url}[1]{#1}
\csname url@samestyle\endcsname
\providecommand{\newblock}{\relax}
\providecommand{\bibinfo}[2]{#2}
\providecommand{\BIBentrySTDinterwordspacing}{\spaceskip=0pt\relax}
\providecommand{\BIBentryALTinterwordstretchfactor}{4}
\providecommand{\BIBentryALTinterwordspacing}{\spaceskip=\fontdimen2\font plus
\BIBentryALTinterwordstretchfactor\fontdimen3\font minus \fontdimen4\font\relax}
\providecommand{\BIBforeignlanguage}[2]{{%
\expandafter\ifx\csname l@#1\endcsname\relax
\typeout{** WARNING: IEEEtranN.bst: No hyphenation pattern has been}%
\typeout{** loaded for the language `#1'. Using the pattern for}%
\typeout{** the default language instead.}%
\else
\language=\csname l@#1\endcsname
\fi
#2}}
\providecommand{\BIBdecl}{\relax}
\BIBdecl

\bibitem[Gonzalez et~al.(2024)Gonzalez, Serna-Torre, S{\'a}nchez-P{\'e}rez, Davidson, Murray, Staadecker, Szinai, Wei, Kammen, Sunter, et~al.]{gonzalez2024offshore}
N.~Gonzalez, P.~Serna-Torre, P.~A. S{\'a}nchez-P{\'e}rez, R.~Davidson, B.~Murray, M.~Staadecker, J.~Szinai, R.~Wei, D.~M. Kammen, D.~A. Sunter \emph{et~al.}, ``Offshore wind and wave energy can reduce total installed capacity required in zero-emissions grids,'' \emph{Nature Communications}, vol.~15, no.~1, p. 6826, 2024.

\bibitem[Harrison-Atlas et~al.(2024)Harrison-Atlas, Glaws, King, and Lantz]{harrison2024artificial}
D.~Harrison-Atlas, A.~Glaws, R.~N. King, and E.~Lantz, ``Artificial intelligence-aided wind plant optimization for nationwide evaluation of land use and economic benefits of wake steering,'' \emph{Nature Energy}, pp. 1--15, 2024.

\bibitem[Zhang et~al.(2014)Zhang, Wang, and Wang]{zhang2014review}
Y.~Zhang, J.~Wang, and X.~Wang, ``Review on probabilistic forecasting of wind power generation,'' \emph{Renewable and Sustainable Energy Reviews}, vol.~32, pp. 255--270, 2014.

\bibitem[Chan and Lee(2012)]{chan2012application}
P.~Chan and Y.~Lee, ``Application of short-range lidar in wind shear alerting,'' \emph{Journal of atmospheric and oceanic technology}, vol.~29, no.~2, pp. 207--220, 2012.

\bibitem[Gao et~al.(2024)Gao, Shen, Wang, Chan, Hon, and Li]{gao2024interpretable}
H.~Gao, C.~Shen, X.~Wang, P.-W. Chan, K.-K. Hon, and J.~Li, ``Interpretable semi-supervised clustering enables universal detection and intensity assessment of diverse aviation hazardous winds,'' \emph{Nature Communications}, vol.~15, no.~1, p. 7347, 2024.

\bibitem[Hon and Chan(2014)]{hon2014application}
K.~Hon and P.~Chan, ``Application of lidar-derived eddy dissipation rate profiles in low-level wind shear and turbulence alerts at h ong k ong i nternational a irport,'' \emph{Meteorological Applications}, vol.~21, no.~1, pp. 74--85, 2014.

\bibitem[Nechaj et~al.(2019)Nechaj, Ga{\'a}l, Bartok, Vorobyeva, Gera, Kelemen, and Polishchuk]{nechaj2019monitoring}
P.~Nechaj, L.~Ga{\'a}l, J.~Bartok, O.~Vorobyeva, M.~Gera, M.~Kelemen, and V.~Polishchuk, ``Monitoring of low-level wind shear by ground-based 3d lidar for increased flight safety, protection of human lives and health,'' \emph{International Journal of Environmental Research and Public Health}, vol.~16, no.~22, p. 4584, 2019.

\bibitem[Dalto et~al.(2015)Dalto, Matu{\v{s}}ko, and Va{\v{s}}ak]{dalto2015deep}
M.~Dalto, J.~Matu{\v{s}}ko, and M.~Va{\v{s}}ak, ``Deep neural networks for ultra-short-term wind forecasting,'' in \emph{2015 IEEE international conference on industrial technology (ICIT)}.\hskip 1em plus 0.5em minus 0.4em\relax IEEE, 2015, pp. 1657--1663.

\bibitem[Liu et~al.(2015)Liu, Tian, and Li]{liu2015emd}
H.~Liu, H.-q. Tian, and Y.-f. Li, ``An emd-recursive arima method to predict wind speed for railway strong wind warning system,'' \emph{Journal of Wind Engineering and Industrial Aerodynamics}, vol. 141, pp. 27--38, 2015.

\bibitem[Klotzbach et~al.(2019)Klotzbach, Blake, Camp, Caron, Chan, Kang, Kuleshov, Lee, Murakami, Saunders, et~al.]{klotzbach2019seasonal}
P.~Klotzbach, E.~Blake, J.~Camp, L.-P. Caron, J.~C. Chan, N.-Y. Kang, Y.~Kuleshov, S.-M. Lee, H.~Murakami, M.~Saunders \emph{et~al.}, ``Seasonal tropical cyclone forecasting,'' \emph{Tropical Cyclone Research and Review}, vol.~8, no.~3, pp. 134--149, 2019.

\bibitem[Zhao et~al.(2023{\natexlab{a}})Zhao, Zhang, Hou, Huang, and Xie]{zhao2023hybrid}
W.~Zhao, Z.~Zhang, B.~Hou, Y.~Huang, and Y.~Xie, ``A hybrid similarity-based method for wind monitoring system deployment optimization along urban railways,'' \emph{Urban Rail Transit}, vol.~9, no.~4, pp. 310--322, 2023.

\bibitem[Abbasi et~al.(2015)Abbasi, Monazzam, Akbarzadeh, Zakerian, and Ebrahimi]{abbasi2015impact}
M.~Abbasi, M.~R. Monazzam, A.~Akbarzadeh, S.~A. Zakerian, and M.~H. Ebrahimi, ``Impact of wind turbine sound on general health, sleep disturbance and annoyance of workers: a pilot-study in manjil wind farm, iran,'' \emph{Journal of Environmental Health Science and Engineering}, vol.~13, pp. 1--9, 2015.

\bibitem[Reitebuch(2012)]{reitebuch2012windlidar}
O.~Reitebuch, ``Wind lidar for atmospheric research,'' in \emph{Atmospheric Physics: Background--Methods--Trends}.\hskip 1em plus 0.5em minus 0.4em\relax Springer, 2012, pp. 487--507.

\bibitem[Rodriguez-Alvarez et~al.(2012)Rodriguez-Alvarez, Akos, Zavorotny, Smith, Camps, and Fairall]{rodriguez2012airborneGNSS}
N.~Rodriguez-Alvarez, D.~M. Akos, V.~U. Zavorotny, J.~A. Smith, A.~Camps, and C.~W. Fairall, ``Airborne gnss-r wind retrievals using delay--doppler maps,'' \emph{IEEE transactions on geoscience and remote sensing}, vol.~51, no.~1, pp. 626--641, 2012.

\bibitem[Liang et~al.(2022)Liang, Wang, Xue, Dou, and Chen]{meter_cdwl}
C.~Liang, C.~Wang, X.~Xue, X.~Dou, and T.~Chen, ``Meter-scale and sub-second-resolution coherent doppler wind lidar and hyperfine wind observation,'' \emph{Optics Letters}, vol.~47, no.~13, pp. 3179--3182, 2022.

\bibitem[Said et~al.(2018)Said, Jelenak, Chang, and Soisuvarn]{said2018assessment}
F.~Said, Z.~Jelenak, P.~S. Chang, and S.~Soisuvarn, ``An assessment of cygnss normalized bistatic radar cross section calibration,'' \emph{IEEE Journal of Selected Topics in Applied Earth Observations and Remote Sensing}, vol.~12, no.~1, pp. 50--65, 2018.

\bibitem[Gleason et~al.(2021)Gleason, Al-Khaldi, Ruf, McKague, Wang, and Russel]{gleason2021characterizing}
S.~Gleason, M.~M. Al-Khaldi, C.~S. Ruf, D.~S. McKague, T.~Wang, and A.~Russel, ``Characterizing and mitigating digital sampling effects on the cygnss level 1 calibration,'' \emph{IEEE Transactions on Geoscience and Remote Sensing}, vol.~60, pp. 1--12, 2021.

\bibitem[Du et~al.(2024)Du, Li, Cardellach, Rib{\'o}, Rius, and Nan]{du2024deep}
H.~Du, W.~Li, E.~Cardellach, S.~Rib{\'o}, A.~Rius, and Y.~Nan, ``Deep residual fully connected network for gnss-r wind speed retrieval and its interpretation,'' \emph{Remote Sensing of Environment}, vol. 313, p. 114375, 2024.

\bibitem[Chu et~al.(2020)Chu, He, Song, Qi, Sun, Bai, Li, and Wu]{chu2020multimodal}
X.~Chu, J.~He, H.~Song, Y.~Qi, Y.~Sun, W.~Bai, W.~Li, and Q.~Wu, ``Multimodal deep learning for heterogeneous gnss-r data fusion and ocean wind speed retrieval,'' \emph{IEEE Journal of Selected Topics in Applied Earth Observations and Remote Sensing}, vol.~13, pp. 5971--5981, 2020.

\bibitem[Reynolds et~al.(2020)Reynolds, Clarizia, and Santi]{reynolds2020wind}
J.~Reynolds, M.~P. Clarizia, and E.~Santi, ``Wind speed estimation from cygnss using artificial neural networks,'' \emph{IEEE Journal of Selected Topics in Applied Earth Observations and Remote Sensing}, vol.~13, pp. 708--716, 2020.

\bibitem[Asgarimehr et~al.(2022)Asgarimehr, Arnold, Weigel, Ruf, and Wickert]{asgarimehr2022gnss}
M.~Asgarimehr, C.~Arnold, T.~Weigel, C.~Ruf, and J.~Wickert, ``Gnss reflectometry global ocean wind speed using deep learning: Development and assessment of cygnssnet,'' \emph{Remote Sensing of Environment}, vol. 269, p. 112801, 2022.

\bibitem[Guo et~al.(2022)Guo, Du, Guo, Southwell, Cheong, and Dempster]{guo2022information}
W.~Guo, H.~Du, C.~Guo, B.~J. Southwell, J.~W. Cheong, and A.~G. Dempster, ``Information fusion for gnss-r wind speed retrieval using statistically modified convolutional neural network,'' \emph{Remote Sensing of Environment}, vol. 272, p. 112934, 2022.

\bibitem[Bu et~al.(2023)Bu, Yu, Zuo, Ni, Li, and Huang]{bu2023glows}
J.~Bu, K.~Yu, X.~Zuo, J.~Ni, Y.~Li, and W.~Huang, ``Glows-net: A deep learning framework for retrieving global sea surface wind speed using spaceborne gnss-r data,'' \emph{Remote sensing}, vol.~15, no.~3, p. 590, 2023.

\bibitem[Zhao et~al.(2023{\natexlab{b}})Zhao, Heidler, Asgarimehr, Arnold, Xiao, Wickert, Zhu, and Mou]{zhao2023ddm}
D.~Zhao, K.~Heidler, M.~Asgarimehr, C.~Arnold, T.~Xiao, J.~Wickert, X.~X. Zhu, and L.~Mou, ``Ddm-former: Transformer networks for gnss reflectometry global ocean wind speed estimation,'' \emph{Remote Sensing of Environment}, vol. 294, p. 113629, 2023.

\bibitem[Clarizia et~al.(2014)Clarizia, Ruf, Jales, and Gommenginger]{clarizia2014spaceborne}
M.~P. Clarizia, C.~S. Ruf, P.~Jales, and C.~Gommenginger, ``Spaceborne gnss-r minimum variance wind speed estimator,'' \emph{IEEE transactions on geoscience and remote sensing}, vol.~52, no.~11, pp. 6829--6843, 2014.

\bibitem[Grey and Gordon(1978)]{grey1978perceptual}
J.~M. Grey and J.~W. Gordon, ``Perceptual effects of spectral modifications on musical timbres,'' \emph{The Journal of the Acoustical Society of America}, vol.~63, no.~5, pp. 1493--1500, 1978.

\bibitem[Hardesty(1986)]{hardesty1986performance}
R.~M. Hardesty, ``Performance of a discrete spectral peak frequency estimator for doppler wind velocity measurements,'' \emph{IEEE transactions on geoscience and remote sensing}, no.~5, pp. 777--783, 1986.

\bibitem[Kliebisch et~al.(2022)Kliebisch, Uittenbosch, Thurn, and Mahnke]{kliebisch2022coherent}
O.~Kliebisch, H.~Uittenbosch, J.~Thurn, and P.~Mahnke, ``Coherent doppler wind lidar with real-time wind processing and low signal-to-noise ratio reconstruction based on a convolutional neural network,'' \emph{Optics Express}, vol.~30, no.~4, pp. 5540--5552, 2022.

\bibitem[Song et~al.(2023)Song, Han, Su, Chen, Sun, Chen, and Xue]{song2023denoising}
Y.~Song, Y.~Han, Z.~Su, C.~Chen, D.~Sun, T.~Chen, and X.~Xue, ``Denoising coherent doppler lidar data based on a u-net convolutional neural network,'' \emph{Applied Optics}, vol.~63, no.~1, pp. 275--282, 2023.

\bibitem[Mohandes and Rehman(2018)]{mohandes2018wind}
M.~A. Mohandes and S.~Rehman, ``Wind speed extrapolation using machine learning methods and lidar measurements,'' \emph{IEEE Access}, vol.~6, pp. 77\,634--77\,642, 2018.

\bibitem[Garc{\'\i}a-Guti{\'e}rrez et~al.(2021)Garc{\'\i}a-Guti{\'e}rrez, Dom{\'\i}nguez, L{\'o}pez, and Gonzalo]{garcia2021atmospheric}
A.~Garc{\'\i}a-Guti{\'e}rrez, D.~Dom{\'\i}nguez, D.~L{\'o}pez, and J.~Gonzalo, ``Atmospheric boundary layer wind profile estimation using neural networks applied to lidar measurements,'' \emph{Sensors}, vol.~21, no.~11, p. 3659, 2021.

\bibitem[Garc{\'\i}a-Guti{\'e}rrez et~al.(2023)Garc{\'\i}a-Guti{\'e}rrez, L{\'o}pez, Dom{\'\i}nguez, and Gonzalo]{garcia2023atmospheric}
A.~Garc{\'\i}a-Guti{\'e}rrez, D.~L{\'o}pez, D.~Dom{\'\i}nguez, and J.~Gonzalo, ``Atmospheric boundary layer wind profile estimation using neural networks, mesoscale models, and lidar measurements,'' \emph{Sensors}, vol.~23, no.~7, p. 3715, 2023.

\bibitem[Gao et~al.(2022{\natexlab{a}})Gao, Shen, Zhou, Wang, Chan, Hon, Zhou, and Li]{gao2022spatio}
H.~Gao, C.~Shen, Y.~Zhou, X.~Wang, P.-W. Chan, K.-K. Hon, D.~Zhou, and J.~Li, ``A spatio-temporal neural network for fine-scale wind field nowcasting based on lidar observation,'' \emph{IEEE Journal of Selected Topics in Applied Earth Observations and Remote Sensing}, vol.~15, pp. 5596--5606, 2022.

\bibitem[Gao et~al.(2022{\natexlab{b}})Gao, Shen, Zhou, Wang, Chan, Hon, and Li]{gao2022deep}
H.~Gao, C.~Shen, Y.~Zhou, X.~Wang, P.-W. Chan, K.-K. Hon, and J.~Li, ``A deep learning-based wind field nowcasting method with extra attention on highly variable events,'' \emph{IEEE Geoscience and Remote Sensing Letters}, vol.~19, pp. 1--5, 2022.

\bibitem[Li et~al.(2024{\natexlab{a}})Li, Huang, Lee, Heo, Ho, and Yim]{li2024integrating}
Y.~Li, T.~Huang, H.~F. Lee, Y.~Heo, K.-F. Ho, and S.~H. Yim, ``Integrating doppler lidar and machine learning into land-use regression model for assessing contribution of vertical atmospheric processes to urban pm2. 5 pollution,'' \emph{Science of The Total Environment}, vol. 952, p. 175632, 2024.

\bibitem[Dosovitskiy(2020)]{dosovitskiy2020image}
A.~Dosovitskiy, ``An image is worth 16x16 words: Transformers for image recognition at scale,'' \emph{arXiv preprint arXiv:2010.11929}, 2020.

\bibitem[Liu et~al.(2024)Liu, Wang, Vaidya, Ruehle, Halverson, Solja{\v{c}}i{\'c}, Hou, and Tegmark]{liu2024kan}
Z.~Liu, Y.~Wang, S.~Vaidya, F.~Ruehle, J.~Halverson, M.~Solja{\v{c}}i{\'c}, T.~Y. Hou, and M.~Tegmark, ``Kan: Kolmogorov-arnold networks,'' \emph{arXiv preprint arXiv:2404.19756}, 2024.

\bibitem[Schmidt-Hieber(2021)]{schmidt2021kolmogorov}
J.~Schmidt-Hieber, ``The kolmogorov--arnold representation theorem revisited,'' \emph{Neural networks}, vol. 137, pp. 119--126, 2021.

\bibitem[Smalikho(2003)]{smalikho2003techniques}
I.~Smalikho, ``Techniques of wind vector estimation from data measured with a scanning coherent doppler lidar,'' \emph{Journal of Atmospheric and Oceanic Technology}, vol.~20, no.~2, pp. 276--291, 2003.

\bibitem[Levin(1965)]{levin1965power}
M.~Levin, ``Power spectrum parameter estimation,'' \emph{IEEE Transactions on Information Theory}, vol.~11, no.~1, pp. 100--107, 1965.

\bibitem[Rui et~al.(2019)Rui, Guo, Chen, Chen, and Zhang]{rui2019adaptive}
X.~Rui, P.~Guo, H.~Chen, S.~Chen, and Y.~Zhang, ``Adaptive iteratively reweighted sine wave fitting method for rapid wind vector estimation of pulsed coherent doppler lidar,'' \emph{Optics Express}, vol.~27, no.~15, pp. 21\,319--21\,334, 2019.

\bibitem[Wei et~al.(2020)Wei, Xia, Wu, Yuan, Wang, and Dou]{wei2020inversion}
T.~Wei, H.~Xia, Y.~Wu, J.~Yuan, C.~Wang, and X.~Dou, ``Inversion probability enhancement of all-fiber cdwl by noise modeling and robust fitting,'' \emph{Optics Express}, vol.~28, no.~20, pp. 29\,662--29\,675, 2020.

\bibitem[Lin et~al.(2021)Lin, Guo, Chen, Chen, and Zhang]{lin2021smoothed}
R.~Lin, P.~Guo, H.~Chen, S.~Chen, and Y.~Zhang, ``Smoothed accumulated spectra based wdswf method for real-time wind vector estimation of pulsed coherent doppler lidar,'' \emph{Optics Express}, vol.~30, no.~1, pp. 180--194, 2021.

\bibitem[Zhang et~al.(2024)Zhang, Zhang, Wang, and Ma]{zhang2024probability}
F.~Zhang, S.~Zhang, L.~Wang, and J.~Ma, ``A probability-constraint-based method based on the honey badger algorithm for wind estimation with coherent doppler wind lidar,'' \emph{Optics Express}, vol.~32, no.~26, pp. 45\,662--45\,678, 2024.

\bibitem[Iandola(2016)]{iandola2016squeezenet}
F.~N. Iandola, ``Squeezenet: Alexnet-level accuracy with 50x fewer parameters and< 0.5 mb model size,'' \emph{arXiv preprint arXiv:1602.07360}, 2016.

\bibitem[Ronneberger et~al.(2015)Ronneberger, Fischer, and Brox]{ronneberger2015unet}
O.~Ronneberger, P.~Fischer, and T.~Brox, ``U-net: Convolutional networks for biomedical image segmentation,'' in \emph{Medical image computing and computer-assisted intervention--MICCAI 2015: 18th international conference, Munich, Germany, October 5-9, 2015, proceedings, part III 18}.\hskip 1em plus 0.5em minus 0.4em\relax Springer, 2015, pp. 234--241.

\bibitem[Henderson et~al.(2005)Henderson, Gatt, Rees, and Huffaker]{henderson2005wind}
S.~W. Henderson, P.~Gatt, D.~Rees, and R.~M. Huffaker, ``Wind lidar,'' in \emph{Laser remote sensing}.\hskip 1em plus 0.5em minus 0.4em\relax CRC Press, 2005, pp. 487--740.

\bibitem[Abdelazim et~al.(2016)Abdelazim, Santoro, Arend, Moshary, and Ahmed]{abdelazim2016signal}
S.~Abdelazim, D.~Santoro, M.~Arend, F.~Moshary, and S.~Ahmed, ``Signal to noise ratio characterization of coherent doppler lidar backscattered signals,'' in \emph{EPJ Web of Conferences}, vol. 119.\hskip 1em plus 0.5em minus 0.4em\relax EDP Sciences, 2016, p. 17014.

\bibitem[Shi et~al.(2024)Shi, Su, Wang, Ni, Duan, and Ren]{shi2024tcnet}
X.~Shi, Q.~Su, W.~Wang, W.~Ni, B.~Duan, and K.~Ren, ``Tcnet: Triple collocation-based network for ocean surface wind speed retrieval on cygnss,'' \emph{IEEE Transactions on Geoscience and Remote Sensing}, 2024.

\bibitem[Vaswani(2017)]{vaswani2017attention}
A.~Vaswani, ``Attention is all you need,'' \emph{Advances in Neural Information Processing Systems}, 2017.

\bibitem[Lei~Ba et~al.(2016)Lei~Ba, Kiros, and Hinton]{lei2016layer}
J.~Lei~Ba, J.~R. Kiros, and G.~E. Hinton, ``Layer normalization,'' \emph{ArXiv e-prints}, pp. arXiv--1607, 2016.

\bibitem[Han et~al.(2024)Han, Zhang, Wu, Zhang, and Wu]{han2024kan4tsf}
X.~Han, X.~Zhang, Y.~Wu, Z.~Zhang, and Z.~Wu, ``Kan4tsf: Are kan and kan-based models effective for time series forecasting?'' \emph{arXiv preprint arXiv:2408.11306}, 2024.

\bibitem[Li et~al.(2024{\natexlab{b}})Li, Liu, Li, Wang, Liu, Liu, Chen, and Yuan]{li2024ukan}
C.~Li, X.~Liu, W.~Li, C.~Wang, H.~Liu, Y.~Liu, Z.~Chen, and Y.~Yuan, ``U-kan makes strong backbone for medical image segmentation and generation,'' \emph{arXiv preprint arXiv:2406.02918}, 2024.

\bibitem[Lhermitte(1961)]{lhermitte1961precipitation}
R.~M. Lhermitte, ``Precipitation motion by pulse doppler.'' in \emph{Proc. 9th Weather Radar Conf., Amer. Meteor. Soc.}, 1961, pp. 218--223.

\bibitem[Loshchilov(2017)]{loshchilov2017decoupled}
I.~Loshchilov, ``Decoupled weight decay regularization,'' \emph{arXiv preprint arXiv:1711.05101}, 2017.

\bibitem[He et~al.(2016)He, Zhang, Ren, and Sun]{hekaiming2016deep}
K.~He, X.~Zhang, S.~Ren, and J.~Sun, ``Deep residual learning for image recognition,'' in \emph{Proceedings of the IEEE conference on computer vision and pattern recognition}, 2016, pp. 770--778.

\bibitem[He et~al.(2022)He, Chen, Xie, Li, Doll{\'a}r, and Girshick]{he2022MAE}
K.~He, X.~Chen, S.~Xie, Y.~Li, P.~Doll{\'a}r, and R.~Girshick, ``Masked autoencoders are scalable vision learners,'' in \emph{Proceedings of the IEEE/CVF conference on computer vision and pattern recognition}, 2022, pp. 16\,000--16\,009.

\end{thebibliography}
\end{document}